# Enhancing Retinal Vascular Structure Segmentation in Images With a Novel Design Two-Path Interactive Fusion Module Model

Rui Yang and Shunpu Zhang

*Department of Statistics and Data Science, University of Central Florida, United States*

**Abstract**

Precision in identifying and differentiating micro and macro blood vessels in the retina is crucial for the diagnosis of retinal diseases, although it poses a significant challenge. Current autoencoding-based segmentation approaches encounter limitations as they are constrained by the encoder and undergo a reduction in resolution during the encoding stage. The inability to recover lost information in the decoding phase further impedes these approaches. Consequently, their capacity to extract the retinal microvascular structure is restricted. To address this issue, we introduce Swin-Res-Net, a specialized module designed to enhance the precision of retinal vessel segmentation. Swin-Res-Net utilizes the Swin transformer which uses shifted windows with displacement for partitioning, to reduce network complexity and accelerate model convergence. Additionally, the model incorporates interactive fusion with a functional module in the Res2Net architecture. The Res2Net leverages multi-scale techniques to enlarge the receptive field of the convolutional kernel, enabling the extraction of additional semantic information from the image. This combination creates a new module that enhances the localization and separation of micro vessels in the retina. To improve the efficiency of processing vascular information, we've added a module to eliminate redundant information between the encoding and decoding steps.

Our proposed architecture produces outstanding results, either meeting or surpassing those of other published models. The AUC reflects significant enhancements, achieving values of 0.9956, 0.9931, and 0.9946 in pixel-wise segmentation of retinal vessels across three widely utilized datasets: CHASE-DB1, DRIVE, and STARE, respectively. Moreover, Swin-Res-Net outperforms alternative architectures, demonstrating superior performance in both IOU and F1 measure metrics.

## 1. Introduction

Retinal examinations can diagnose many retinal conditions, such as diabetic retinopathy, epiretinal membrane, macular edema, and cytomegalovirus retinitis. Retinal vascular disorders, which attack the retinal blood vessels, are typically connected to other diseases such as atherosclerosis, hypertension or changes in the circulatory area [1, 2], and the precise segment of retinal blood vessels and the identification of the playful space of retinal area disorders are needed to determine their pertinent diagnosis.

In recent years, retinal vessel segment methods have been proposed based on the methods of image processing and machine learning [3-6]. In particular, although the detection performance of the vessel appears to be generally improved, it is difficult to obtain pixel-accurate segmentation in some cases because of factors such as insufficient illumination of the image and periodic noise, likely resulting in a large number of false positives [3].

Early research on vessel segmentation primarily focused on techniques using hand-crafted features [7, 8], filter-based models [9], and statistical models [10]. By improving border gradients, eliminating irrelevant background information, and filtering picture noise, these techniques aim to simplify the segmentation problem to a mathematical optimization problem with a predetermined solution. The advancements of data-driven methodologies and technology in computers have made deep learning an important field of research and application in medical image analysis. Deep learning has been extensively studied for its remarkable representational learning capabilities [11], consistently outperforming traditional data segmentation approaches.

From 2012 to 2020, convolutional neural networks (CNNs) dominated the integration of medical imaging with deep learning. CNNs extract shallow and deep visual features layer by layer by using multiple convolutional layers, pooling layers, and fully connected layers at the bottleneck. For medical image segmentation, models like Deeplab-v3 [12] have not been widely adopted due to the complexity of the models and the features of medical images. Currently, the basic model for deep learning-based medical image segmentation is the U-Net model [13] proposed in 2015. Future advancements in this field are based on its symmetric encoder and decoder structure.

Although CNNs have good feature extraction ability, their inherent inductive bias property limits their attention to local image features, hindering further model performance improvement. Some works have introduced CNN-based network attention mechanisms, such as Compression Excitation Networks [14] and Axial DeepLab [15]. However, these studies have not significantly addressed the natural deficiencies of CNNs.

The Deep Self-Attention Network (Transformer), first proposed in the article "Attention is All You Need" [16], originally used in natural language processing, has become the cornerstone of large-scale models such as GPT-3. The Transformer's ability to model long-distance correlations and focus on the global properties of input information makes it ideal for areas such as language translation. Since 2020, researchers have explored applying Transformers to computer vision with significant progress. Google's ViT [17], Facebook's DeiT [18], and Microsoft Research Asia's Swin Transformer [19] are excellent examples. Swin Transformer has been widely used in various computer vision tasks, including medical image segmentation [19-21], showing its potential to match or exceed CNN performance, opening new avenues for computer vision development.

In this paper, we propose a novel model that integrates the Swin Transformer into U-Net, improving the network's capacity to capture long-distance dependencies and model global information. This addresses the limitations of convolutional networks, which predominantly focus on local details, resulting in a more precise segmentation of small vessels. In theory, increasing network depth should improve performance. In practice, deeper networks pose challenges for training optimization algorithms, leading to increased training errors. To address issues like gradient vanishing, our model utilizes Res2net, ensuring that training efficiency is maintained even with a deeper network. This novel technique uses multi-scale approaches to increase the convolutional kernel's receptive field, which enables the extraction of more semantic information from the image. Additionally, our model includes a mechanism to remove redundant information between the encoder and decoder, enhancing overall model efficiency.

## 2. Methodology

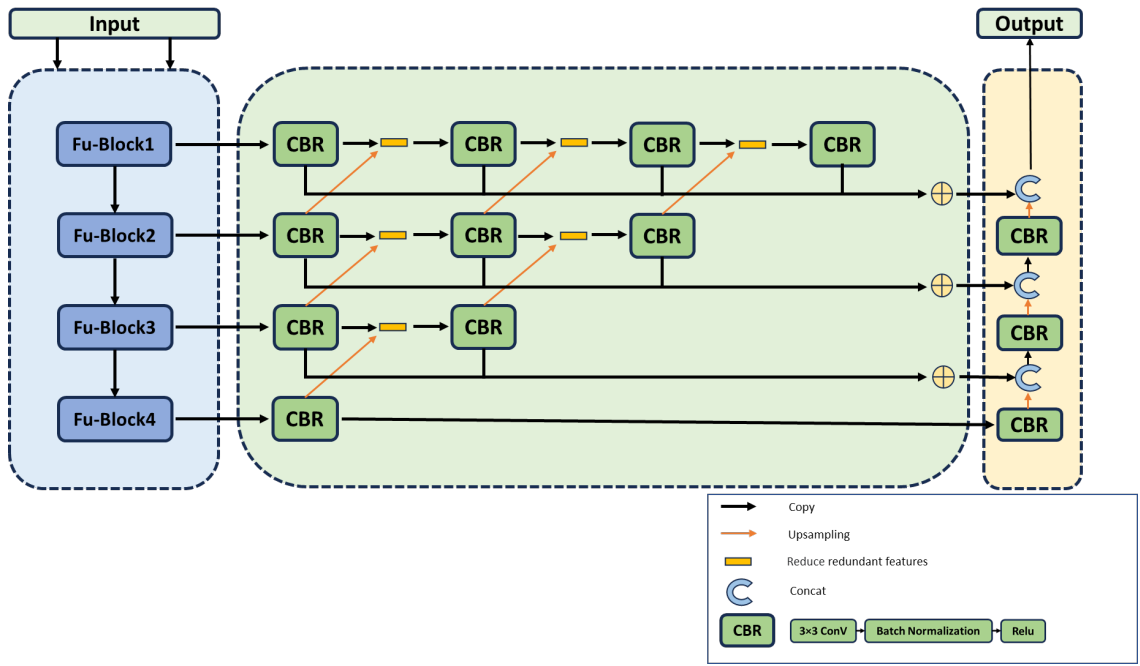

Figure 1: Architecture of the model structure. It consists of three parts: a feature extraction encoder (light blue), a redundant information reduction module (light green) and a decoder (light orange).

### 2.1 Overall of the model

The model proposed in this paper is based on the U-Net neural network architecture. The traditional U-Net structure follows a U-shaped design, consisting of three key components: an encoder, a decoder, and connections. As illustrated in Figure 1, the encoder of our proposed model comprises four groups of Fu_Block modules, utilizing convolution and downsampling techniques to extract context information from the feature map. In contrast, the decoder includes three groups of CBR modules and up-sampling modules to restore the resolution of the feature map. The CBR module consists of a 3 × 3 convolution layer, a batch normalization layer, and a ReLU activation layer. Finally, to ensure effective communication between the encoder and decoder levels, a redundant information reduction module is employed. This module is crucial because multiple convolutions may cause the feature map to lose spatial information. This is in contrast to the U-Net model, where there is no module for reducing redundant information, in order to easily affect the combination of the context-rich feature map from the encoder with the feature map from the decoder and at the same time inhibit excess transfer of information to the decoder, thus increasing the efficiency of the model.

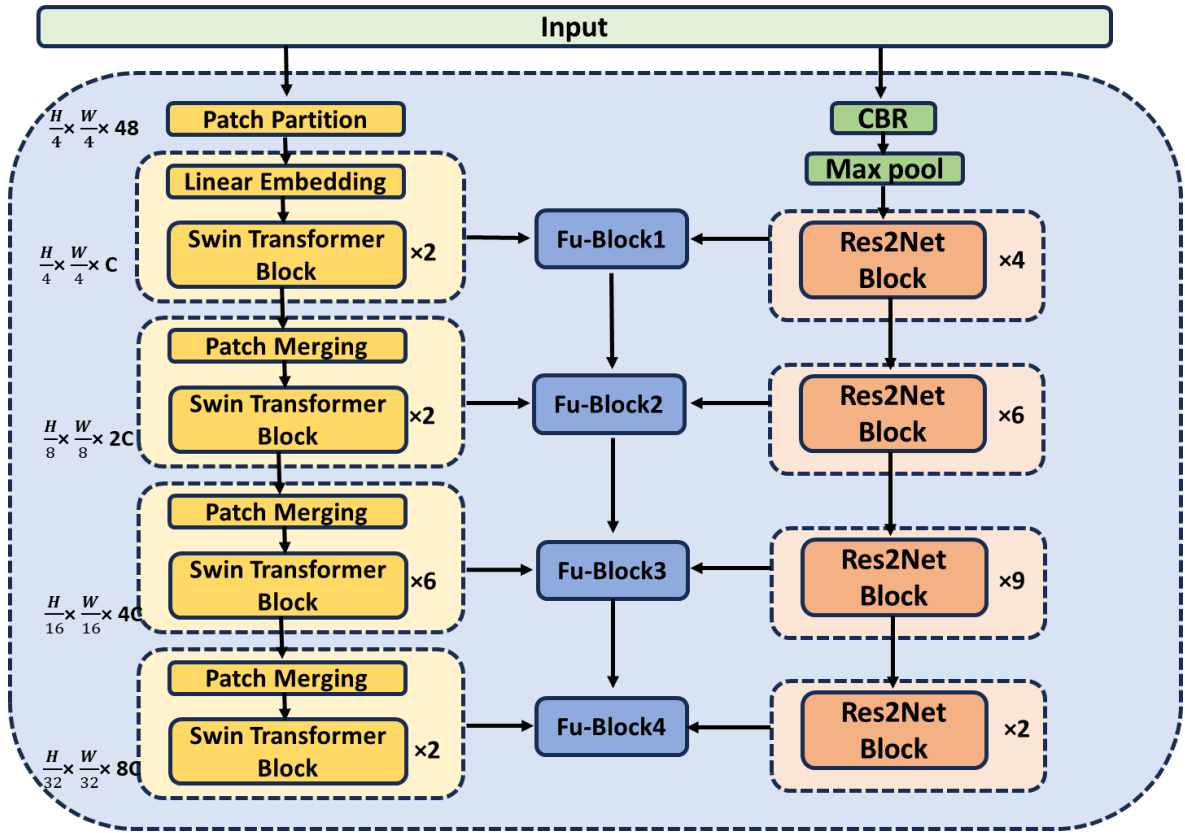

Figure 2: Feature extraction encoder

## 2.2 Encoder

Figure 2 illustrates the detailed structure of the feature extraction encoder. The preprocessed image gets into the Swin Transformer method and the Residual Block method in the encoder. On the other hand, in both paths of the four-layer, the output of each layer has been made by the two methods. Thereafter, the two layers are converged, and the fusions sent to the redundant information reduction module.

### 2.2.1 Swin Transformer path

Inspired by its remarkable achievements in segmentation tasks, we have incorporated the Swin Transformer into the U-Net architecture. With this integration, the network can recognize the long-range dependencies and therefore be capable of establishing a coherent understanding of the global context. This development effectively overcomes the limitation of the classical convolutional network, which is only programmed to process local information. Consequently, our method improves the accuracy of identifying and segmenting fine details, like small blood vessels.

The initial step involves dividing a preprocessed RGB image into non-overlapping patches using a patch splitting module. Each patch is considered a "token," with its feature represented as the concatenation of raw pixel RGB values. For our work, we use a patch size of 4 × 4, which means the feature dimension will be 4 × 4 × 3 = 48. We then apply a linear embedding layer to this raw-valued feature, which projects it to an arbitrary dimension, denoted as C.

Subsequently, multiple Transformer blocks, incorporating customized self-attention computations and referred to as Swin Transformer blocks, are applied to these patch tokens. These Transformer blocks maintain the token count at H/4 × W/4, forming what we term 'Stage 1.'

To generate a hierarchical representation, the reduction in the number of tokens occurs through patch merging layers as the network deepens. The first layer merges these two concatenated patch features into one by concatenating the features of each pair of 2×2 neighboring patches and applying a linear layer to these concatenated 4C-dimensional features. This process results in a fourfold reduction in tokens (equivalent to a 2×2 down-sampling of resolution), while adjusting the output dimension to 2C. Following this, Swin Transformer blocks perform feature transformation while maintaining the resolution at H/8 × W/8. The procedure progresses through four phases, with notable developments occurring in 'Stage 3' and 'Stage 4.' Here, the resolution of the output is enhanced to H/16 × W/16 and to H/32 × W/32, respectively.

The Swin Transformer is constructed by replacing the standard multi-head self-attention (MSA) module in a Transformer block with a module based on shifted windows, while keeping other layers unchanged. As depicted in Figure 3 (a), a Swin Transformer block is basically a shifted window-based MSA module, followed by a 2-layer MLP with GELU nonlinearity. A LayerNorm (LN) layer is applied before each MSA module and each MLP, and a residual connection is applied after each module. The calculation process of a continuous Swin Transformer block can be expressed by a series of formulas:

$$\hat{z}^l = W\text{-}MSA(\text{LN}(z^{l-1})) + z^{l-1},$$

$$z^l = MLP(LN(\hat{z}^l)) + \hat{z}^l,$$

$$\hat{z}^{l+1} = SW\text{-}MSA(\text{LN}(z^l)) + z^l,$$

$$z^{l+1} = MLP(LN(\hat{z}^{l+1})) + \hat{z}^{l+1}, \tag{1}$$

Here $z^l$ denotes the output features of the SW-MSA module and the MLP module for block $l$, respectively. W-MSA and SW-MSA represent window-based multi-head self-attention using regular and shifted window partitioning configurations, respectively.

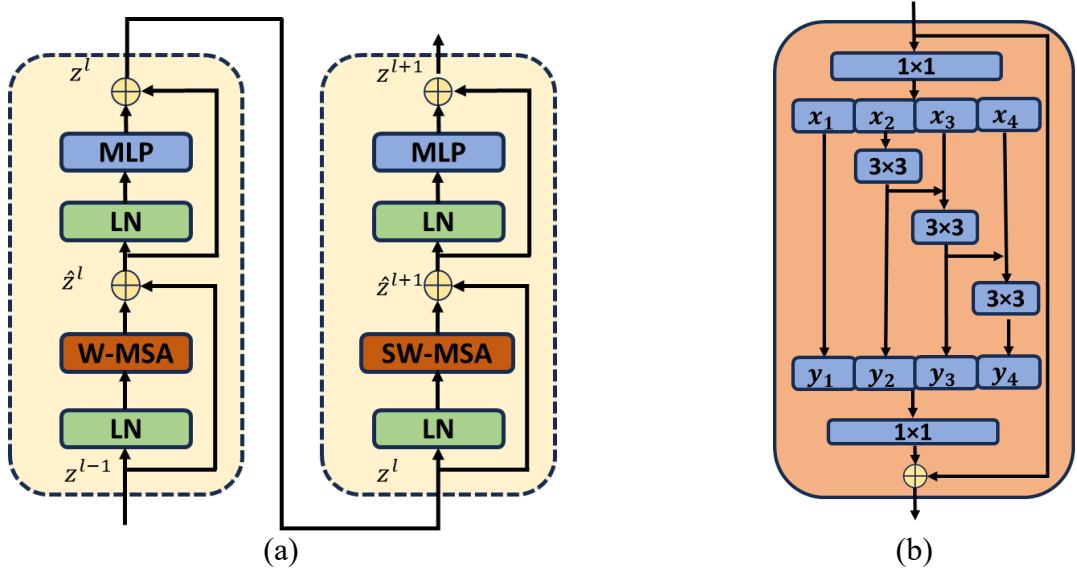

Figure 3: (a) The basic structure of consecutive Swin Transformer blocks; (b) The basic structure of residual blocks (Res2 module).

### 2.2.2 Residual Block path

A Res2Net block has been introduced as an additional module in another path. In the segmentation of fundus image vessels, the distribution of micro-vessels is diffuse, and their size is small. Hence, there is a high probability of loss of semantic information on small objects with a number of single convolution operations that lead to the segmentation accuracy of the small objects being comparatively reduced [22]. Drawing inspiration from Res2Net, this paper introduces and incorporates the Res2Net Block into U-Net. This consists of partitioning the feature map across numerous channels, merging the adjacent feature maps, and later applying convolution in order to develop the receptive field of the network. This new method is designed to overcome obstacles in accurately segmenting small objects, focusing on improving the retrieval of semantic information within the process of segmenting vessels in fundus images.

The process begins with the input of a preprocessed RGB image into the CBR module and max pool. Subsequently, we apply Res2Net blocks 4, 6, 9, and 2 times across the four layers. Figure 3 (b) displays the configuration of the Res2Net block, beginning with the application of a $1 \times 1$ convolution kernel to the input feature map, followed by the division of the channel into four separate groups. While the first group is directly transmitted downward, all other groups have a $3 \times 3$ convolution kernel, which is used to extract features and bring about a change in the receptive field in the branch. Each group's output, fused with the feature map on the left, is applied. Group splicing and fusion are then done through a $1 \times 1$ convolution kernel. In the final step, the outcome is combined with the output from the residual connection branch. This approach is based on multi-scale methods that increase the receptive field of the convolutional kernel, enabling further semantic information extraction from images.

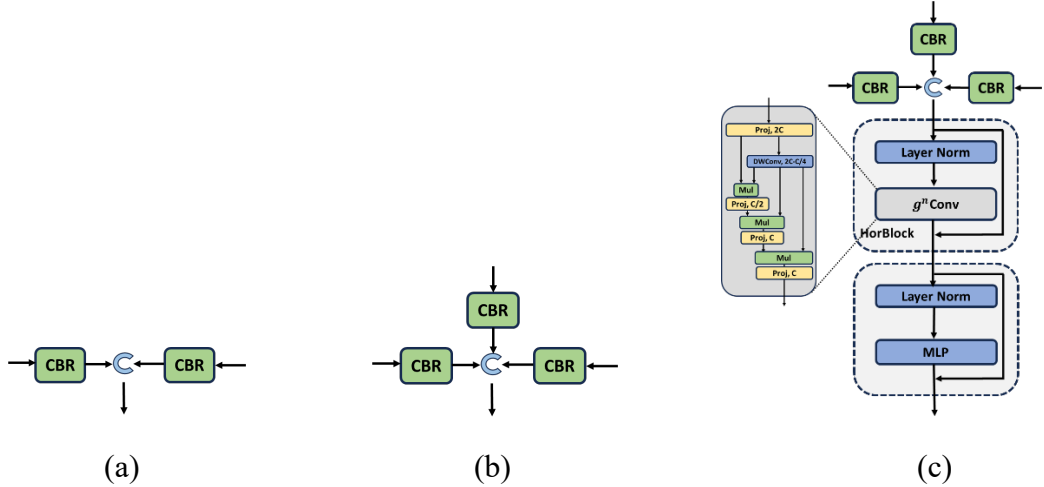

Figure 4: The basic structure of fusion blocks. (a) Fu-Block 1; (b) Fu-Block 2 and 3; (c) Fu-Block 4

### 2.2.3 Fusion of the outputs of two paths

As depicted in Figure 2, the two paths are fused separately at the output of each layer. Figure 4 illustrates the structure of four Fu-Blocks, which perform CBR on each input and then utilize concatenation to connect the 1-dimensional feature matrix. Fu-Block 1 has two inputs: one from the output of the 1st layer Swin Transformer block and the other from the 1st layer Residual block. Fu-Blocks 2, 3, and 4 have three inputs, two of which are identical to those in Fu-Block 1 from the Swin Transformer Block and Residual Block, while the remaining input is copied from the preceding Fu-Block.

The output of Fu-Block 4 distinguishes itself from the others through the implementation of fusion coding using a Horblock. This specialized approach is employed to capture attention and enhance fusion attention features. Notably, Horblock incorporates Recursive Gated Convolution ($g^nConv$), as illustrated in Figure 4 (c). In this newly designed approach, it has been intended to support the efficient, translation-equivariant, and extendable high-order spatial interactions via a recursive architecture with the help of gated convolutions.

It's important to emphasize that $g^nConv$ can seamlessly replace the spatial mixing layer in various Vision Transformers and convolution-based models. The quadratic complexity in input size related to self-attention encumbers the limit for how practically it can be efficiently used within Vision Transformers, more so in tasks like segmentation and detection, which require higher resolution for its feature maps.

### 2.3 Reduce redundant information module.

In Figure 1, the light green block represents the module to reduce the redundant information. It is responsible for providing the capacity to combine the context-rich feature map from the encoder with the feature map from the decoder. It also prevents oversupply of information to the decoder, thus making this overall model more efficient.

Bilinear interpolation is employed as the method for upsampling, effectively enlarging the features in layer n+1 to match the spatial dimensions of layer n. After the upsampling, the module measures the absolute element-wise difference between the upsampled tensor of the current layer n+1 and the tensor of the previous layer n. This computation quantifies the absolute difference between the two feature maps, preserving the most important information and enabling copy and crop to the decoder. The functionality of the following code will depend on the characteristic and the architecture of the deep learning model to apply. The module can be expressed using the formulas provided below.

$$f(x)^{1'} = abs(CBR(f(x)^1) - \uparrow[CBR(f(x)^2)])$$

$$f(x)^{1''} = abs(CBR(f(x)^{1'}) - \uparrow[CBR(f(x)^{2'})])$$
$$\vdots$$
$$f(x)^{2''} = abs(CBR(f(x)^{2'}) - \uparrow[CBR(f(x)^{3'})]) \qquad (2)$$
$$\vdots$$

Where $f(x)$ is the feature map for each layer. The number of superscripts is the layer number. The prime symbol is the number of reduced redundant operations. ↑ represents the bilinear interpolation unsampling method.

## 2.4 Decoder

In Figure 1, the light-yellow section represents the decoder component of our model. At Layer 4, the input undergoes a single cycle of CBR (convolution, batch normalization, and ReLU activation) before being unsampled and concatenated with the input at Layer 3. This process repeats until the final output is obtained.

## 3. Experiments

### 3.1 Dataset

The DRIVE [23] dataset is a collection of 40 retinal images with segmentation annotations, obtained from a diabetic retinopathy screening program in the Netherlands. Seven of the images show mild early diabetic retinopathy, and 33 show normal ones. The retinal images were formally divided into training sets. The images are captured at a size of 565 × 584 pixels with 8 bits per color plane.

The CHASE_DB1 [24] dataset, designed for retinal vascular segmentation, contains 28 color retina images of 999 × 960 pixels' resolution, taken from the left and right eyes of 14 students. All the sets of these images were manually annotated for segmentation by two independent experts, normally taking the annotations of the first expert as the ground truth. The first 20 images are meant for training, while the remaining eight are reserved for testing.

The STARE [25] dataset encompasses 20 retinal fundus images, half of which exhibit pathological signs. During the iteration, 18 images are selected at a time and are considered as training samples, while the remaining set of images will be used as test samples over 10 repetitions. There was no predefined division of the data to solidify the reliability of the experimental results, hence making 10-fold cross-validation a suitable method to apply.

### 3.2 Preprocessing

Image pre-processing was performed with the aim of increasing the data diversity and therefore making the model more robust before entering the model. (1) A random horizontal flip operation with a 0.5 probability was applied. (2) A random vertical flip with a 0.5 probability was employed. (3) Random rotations were applied to the images.

### 3.3 Hyper-parameter

Hyperparameter settings used in our model training are as follows. Optimizer: Adam. Initial learning rate: $10^{-4}$. Weight decay: $10^{-5}$. Learning rate adjustment strategy: CosineAnnealingLR. Post-processing image threshold: 0.5. Number of training epochs: 40.

### 3.4 Loss function: Binary Cross-Entropy Loss (BCELoss).

The model employs a binary cross-entropy loss with a fixed threshold of 0.5 to determine whether a pixel belongs to a vessel or the background. The unreduced loss can be described as:

$$\mathcal{L}(y, \hat{y}) = -\frac{1}{N}\sum_{i=1}^{N}[y_i \log \hat{y}_i + (1 - y_i)\log(1 - \hat{y}_i)] \quad (3)$$

where $y$ and $\hat{y}$ indicate ground truth and predicated of $i^{th}$ image, N is the batch size.

### 3.5 Quantitative Benchmarking

We conducted an extensive comparative analysis, assessing our architecture alongside several high-performing models, including Unet++ [26], CS-Net [27], Residual U-Net [28], RV-GAN [29], and FR-Unet [30]. Training and evaluation were performed using their publicly accessible source code on all three datasets.

Subsequently, we have compared our architecture with the existing retinal vessel segmentation models and presented the results for datasets DRIVE, CHASE-DB1, and STARE in Table 1. Sensitivity, specificity, accuracy, F1-score, and area under the curve (AUC) are some of the conventional performance evaluation parameters that are calculated. We have further investigated the retinal vessel segmentation accuracy and structural similarity using intersection-over-union (IOU). These evaluations offer a comprehensive understanding of the strengths and weaknesses of each model in our comparative analysis.

As observed in the tables, our model is ranked consistently with better performance compared with the existing U-Net based design and even the more recent GAN-based models. These are evidenced by the AUC, F1 score, and IOU, which are the main evaluation metrics in this task. Notably, our model improved specificity, accuracy, AUC, F1 score, and IOU values across all datasets.

| Method | Year | Sensitivity | Specificity | Accuracy | AUC | F1 | IOU |
|---|---|---|---|---|---|---|---|
| Unet++ | 2018 | 0.8357 | 0.9832 | 0.9739 | 0.9881 | 0.7898 | 0.6526 |
| CS-Net | 2020 | 0.8400 | 0.9832 | 0.9742 | 0.9881 | 0.8042 | 0.6725 |
| Residual U-Net | 2021 | 0.8178 | 0.9822 | 0.9644 | 0.9834 | - | - |
| RV-GAN | 2021 | 0.8199 | 0.9806 | 0.9697 | 0.9914 | **0.8957** | - |
| FR-Unet | 2022 | **0.8798** | 0.9814 | 0.9748 | 0.9913 | 0.8151 | 0.6882 |
| Swin-Res-Net | 2024 | 0.8467 | **0.9918** | **0.9813** | **0.9956** | 0.8665 | **0.7646** |

Table 1 (a): Result of vessel segmentation (CHASE_DB1)

| Method | Year | Sensitivity | Specificity | Accuracy | AUC | F1 | IOU |
|---|---|---|---|---|---|---|---|
| Unet++ | 2018 | 0.7891 | 0.9850 | 0.9679 | 0.9825 | 0.8114 | 0.6827 |
| CS-Net | 2019 | 0.8170 | 0.9854 | 0.9632 | 0.9798 | 0.8039 | 0.7017 |
| Residual U-Net | 2021 | 0.8062 | 0.9825 | 0.9683 | 0.9802 | - | - |
| RV-GAN | 2021 | 0.7927 | **0.9969** | **0.9790** | 0.9887 | **0.8690** | - |
| FR-Unet | 2022 | **0.8356** | 0.9837 | 0.9705 | 0.9889 | 0.8316 | 0.7120 |
| Swin-Res-Net | 2024 | 0.8214 | 0.9872 | 0.9728 | **0.9931** | 0.8394 | **0.7234** |

Table 1 (b): Result of vessel segmentation (Drive dataset)

| Method | Year | Sensitivity | Specificity | Accuracy | AUC | F1 | IOU |
|---|---|---|---|---|---|---|---|
| Unet++ | 2018 | 0.7909 | 0.9883 | 0.9734 | 0.9884 | 0.8118 | 0.6856 |
| CS-Net | 2020 | 0.7926 | 0.9882 | 0.9735 | 0.9885 | 0.8159 | 0.6912 |
| Residual U-Net | 2021 | 0.7930 | 0.9883 | 0.9716 | 0.9869 | - | - |
| RV-GAN | 2020 | 0.8356 | 0.9864 | 0.9754 | 0.9887 | 0.8323 | - |
| FR-Unet | 2022 | 0.8327 | 0.9869 | 0.9752 | 0.9914 | 0.8330 | 0.7156 |
| Swin-Res-Net | 2024 | **0.8383** | **0.9892** | **0.9779** | **0.9946** | **0.8498** | **0.7389** |

Table 1 (c): Result of vessel segmentation (STARE Dataset) data and charts.

### 3.6 Precision of vessel segmentations

As illustrated in Figure 5, Swin-Res-Net demonstrates a high level of accuracy in vessel segmentation compared to ground truths. The most challenging task is micro blood vessel segmentation. Notably, our model successfully identified and segmented micro blood vessels.

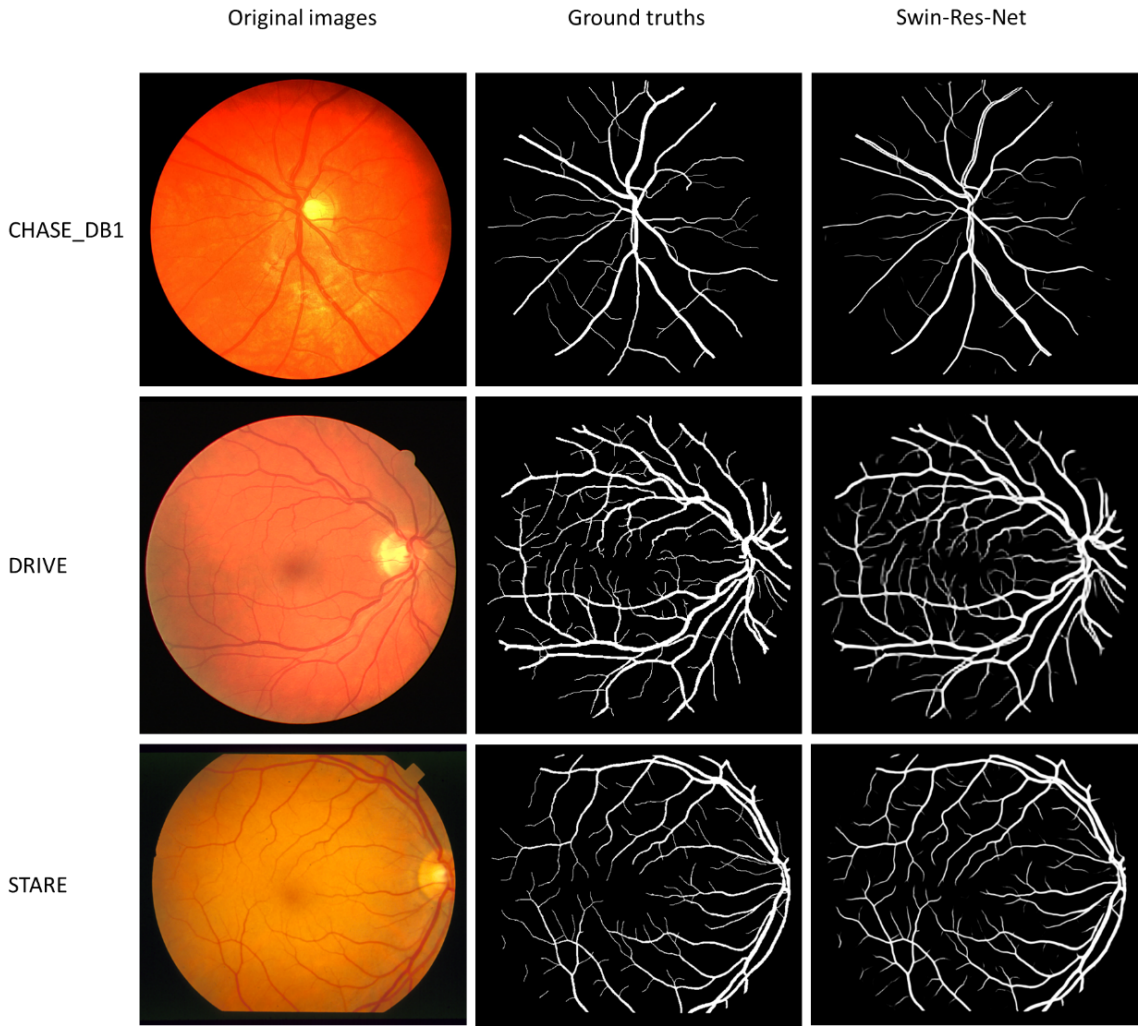

Figure 5: Swin-Res-Net segments vessel with good precision on the micro blood vessel

## 4.      Conclusions

In this research paper, we introduce Swin-Res-Net, a new multiscale architecture. By integrating innovative matching loss functionality, this architecture demonstrates the ability to generate highly accurate segmentations of vein structures while providing robust confidence values for two key performance parameters. Our architectural innovation holds significant application potential in the field of ophthalmology, particularly in analyzing degenerative retinal diseases and predicting future developments. Our future research efforts aim to extend the application of this methodology to diverse data modalities.